\documentclass[sigconf]{acmart}
\AtBeginDocument{%
  \providecommand\BibTeX{{%
    \normalfont B\kern-0.5em{\scshape i\kern-0.25em b}\kern-0.8em\TeX}}}

\usepackage{mathtools}
\usepackage{tabularx}
\usepackage{tablefootnote}
\usepackage{tikz}
\usepackage[shortcuts,acronym]{glossaries}
\usepackage{titlesec}
\usepackage{dblfloatfix}
\usetikzlibrary{mindmap}

\makeglossaries 
\setacronymstyle{long-short}
\newacronym{gan}{GANs}{Generative Adversarial Networks}
\newacronym{pii}{PII}{Personally Identifiable Information}
\newacronym{pcfg}{PCFG}{Probabilistic Context-Free Grammar}
\newacronym{rnn}{RNN}{Recurrent Neural Networks}
\newacronym{lstm}{LSTM}{Long Short Term Memory Networks}
\newacronym{cnn}{CNN}{Convolutional Neural Networks}
\newacronym{psm}{PSMs}{Password Strength Meters}
\newacronym{wae}{WAEs}{Wasserstein Autencoders}
\newacronym{dpg}{DPG}{Dynamic Password Guessing}
\newacronym{bert}{BERT}{Bidirectional Encoder Representations from Transformers}
\newacronym{bilstm}{BiLSTM}{Bidirectional Long Short-term Memory}
\newacronym{fla}{FLA}{Fast, Lean, Accurate}
\newacronym{iwgan}{IWGAN}{Improved Training of Wasserstein GANs}
\newacronym{sprnn}{SPRNN}{Structure Partition and Bilstm Recurrent Neural Network}
\begin{document}

\settopmatter{printfolios=true}
\title{On Deep Learning in Password Guessing, a Survey}

\author{Fangyi Yu}
\email{fangyi.yu@ontariotechu.net}
\affiliation{%
  \institution{Ontario Tech University}
  \streetaddress{2000 Simcoe St N}
  \city{Oshawa}
  \state{Ontario}
  \country{Canada}
  \postcode{L1G 0C5}
}


\begin{abstract}
The security of passwords is dependent on a thorough understanding of the strategies used by attackers. Unfortunately, real-world adversaries use pragmatic guessing tactics like dictionary attacks, which are difficult to simulate in password security research. Dictionary attacks must be carefully configured and modified to be representative of the actual threat. This approach, however, needs domain-specific knowledge and expertise that are difficult to duplicate. This paper compares various deep learning-based password guessing approaches that do not require domain knowledge or assumptions about users' password structures and combinations. The involved model categories are \ac{rnn}, \ac{gan}, autoencoder and attention mechanisms. Additionally, we proposed a promising research experimental design on using variations of \ac{iwgan} on password guessing under non-targeted offline attacks. Using these advanced strategies, we can enhance password security and create more accurate and efficient \ac{psm}.

\end{abstract}


\maketitle

\section{Introduction}

Passwords have dominated the authentication system for decades, despite their security flaws compared to competing techniques such as cognitive authentication and hardware tokens. The irreplaceable is primarily due to its incomparable deployability and usability\cite{bonneau2012quest}. However, the security of user-selected passwords continues to be a significant concern. According to research examining susceptible behaviours that affect password crackability\cite{adams1999users}, there are three types of user actions that result in the creation of insecure passwords: 1. Users often use basic terms in passwords and perform simple string transformations to comply with websites' password creation policies\cite{narayanan2005fast}. 2. Password reuse is prevalent since the typical user has more than 20 accounts, and establishing unique passwords and remembering them for each account is too time-consuming\cite{stobert2014password}. According to S.Pearman et al., 40\% of users repeat their passwords\cite{pearman2017let}. 3. Users prefer to use simple-to-remember passwords that include personal information such as their birth date and their pets' name. All of these behaviours expose the user-created passwords to attacks. Additionally, the recent large-scale leakage of passwords on multiple platforms across the world (listed in Table\ref{tab:datasets}) raises the alarm for researchers.  

\subsection{Offline attacks and Online attacks}
Password guessing attacks fall into two categories: offline and online. Offline attacks occur when attackers get cryptographic hashes of certain users' passwords and attempt to recover them by guessing and testing many passwords. The primary objective is to determine the difficulty of cracking a genuine user's password, or the strength of a user-created password, by producing a list of password guesses and checking for the possibility of the genuine user's password's occurrence. Offline attacks are only considered when the following conditions are met: An attack gains access to the system and extracts the password file, all while remaining unnoticed. Moreover, the file's salting and hashing must be done appropriately. Otherwise, an offline assault is either ineffective (the attacker may get credentials directly without requiring guesses, or an online approach is more effective), or impossible\cite{florencio2014administrator}. 

An online attack occurs when an attacker makes password attempts against users using a web interface or an application. This situation is more constrained for attackers since most authentication systems automatically freeze accounts after several unsuccessful attempts. Therefore, attackers must guess users' passwords successfully within the allotted number of tries, which is the primary difficulty of online password guessing. According to Florencio et al. \cite{florencio2014administrator}, $10^6$ is a reasonable upper limit for the number of online guesses a secure password must survive, while the number of offline guesses is difficult to quantify considering the attacker's possible usage of unlimited computers each calculating hashes thousands of times quicker than the target site's backend server.

\subsection{Targeted attacks and non-targeted attacks}
Targeted guessing attacks occur when attackers attempt to break users' passwords using their knowledge on users, specifically their \ac{pii}, such as name, birth date, anniversary, home address etc,. This is a considerable concern when \ac{pii} becomes more accessible as a result of constant data breaches. On the contrary, non-targeted attacks do not assume the users' identities.
Be aware that the majority of individual accounts are not deserving of concentrated attention. Only significant or vital accounts relating to critical work, finances, or documents demanding a high level of security may be considered\cite{florencio2014administrator}.

This paper primarily focus on utilizing deep learning to increase passwords guessability under the offline and non-targeted scenario.

\subsection{Contributions}

The following are the primary contributions of this paper: 
\begin{itemize}
\item  It provides an in-depth comparison of deep learning techniques used for untargeted offline password guessing.  
\item  It identifies the open challenges and possible directions for future study in this topic. 
\item  It proposes a viable research approach for password guessing using \ac{gan} and a feasible experimental design. 
\end{itemize}

The rest of the paper is divided into the following sections: Section 2 discusses three approaches for password guessing: rule-based, probability-based, and deep learning-based models. The third section discusses methodology, datasets, and experimental design, and section 4 discusses unresolved issues and future work. The last part contains abbreviations and a bibliography.

\section{Background and Related Work}
The three predominant ways of password guessing are rule-based, probability-based, and deep learning-based. 

\subsection{Rule-based models}
The large amount of stolen passwords simplifies the process of collecting password patterns. Following that, other candidate passwords may be produced using these password patterns as guidelines. Hashcat\footnote{https://hashcat.net/wiki/} and John the Ripper \footnote{https://www.openwall.com/john/}are two popular open-source password guessing programs that use rule-based password guessing. They provide a variety of ways for cracking passwords, including dictionary attacks, brute-force attacks, and rule-based attacks. Among all these types, the rule-based one is the fastest, and HashCat is the market leader in terms of speed, hash function compatibility, updates, and community support\cite{hranicky2019distributed}. However, rule-based systems create passwords solely based on pre-existing rules, and developing rules requires domain expertise. Once rules are defined, passwords that violate those restrictions will not be identified.

\subsection{Probability-base models}
Apart from rule-based password guessing models, conventional password guessing models are mostly probability-based, with two notable approaches being Markov Models and \ac{pcfg}. Markov Models are built on the assumption that all critical password features can be specified in n-grams. Its central principle is to predict the next character based on the preceding character\cite{narayanan2005fast}. \ac{pcfg} examines the grammatical structures (combinations of special characters, digits, and alphanumerical sequences) in disclosed passwords and generates the distribution probability, after which it uses the distribution probability to produce password candidates\cite{weir2009password}.

\subsection{Deep Learning-base models}
Unlike rule-based or probability-based password guessing tools, deep learning-based methods make no assumptions about the password structure. Deep neural network-generated password samples are not constrained to a particular subset of the password space. Rather than that, neural networks can autonomously encode a broad range of password information beyond the capabilities of human-generated rules and Markovian password-generating methods. 

\subsubsection{Recurrent Neural Networks}

\begin{figure}[t]
  \includegraphics[width=0.45\textwidth]{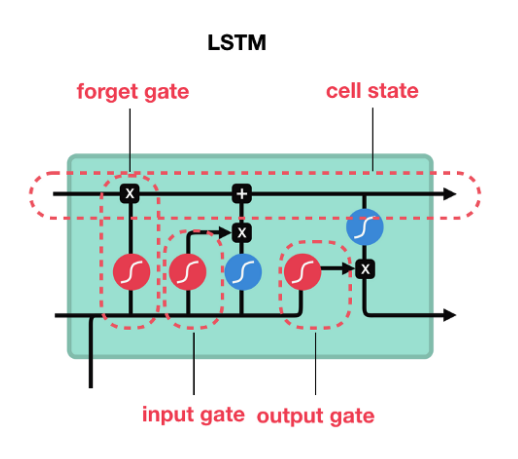} 
  \caption{The Gating Mechanism in \ac{lstm}.}
  \label{fig:LSTM}
\end{figure}

\ac{rnn} are neural networks in which inputs are processed sequentially and restored using internal memory. They are often employed to solve sequential tasks such as language translation, natural language processing, voice recognition, and image recognition. Due to the fact that the vanilla \ac{rnn} architecture is incapable of processing long-term dependencies due to the vanishing gradient issue\cite{pascanu2013difficulty}, therefore, \ac{lstm} was designed to tackle the problem. The \ac{lstm} makes use of the gating mechanism depicted in Figure\ref{fig:LSTM} to retain information in memory for long periods of time\cite{hochreiter1997long}.

To the best of our knowledge, Melicher et al.\cite{melicher2016fast} were the first to utilize \ac{rnn} to extract and predict password features. They kept their model, named \ac{fla}, as lightweight as possible in order to integrate it into local browsers for proactive password verification. Three \ac{lstm} layers and two highly connected layers comprise the proposed Neural Network. Various strategies for training neural networks on passwords were used. It was proven that employing transfer learning\cite{yosinski2014transferable} significantly improves guessing efficacy, however, adding natural language dictionaries to the training set and tutoring had little impact. Consequently, they discovered that Neural Networks are superior at guessing passwords when the number of guesses is increased and when more complicated or longer password policies are targeted. Nevertheless, because of the Markovian nature of \ac{fla}'s password generation process, any password feature that is not included within the scope of an n-gram may be omitted from encoding\cite{hitaj2019passgan}.

Zhang et al.\cite{zhang2018password} presented \ac{sprnn}, a hybrid password attack technique based on structural partitioning and \ac{bilstm}. The \ac{pcfg} is used for structure partitioning, which seeks to structure the password training set to learn users' habit of password construction and generate a collection of basic structures and string dictionaries ordered by likelihood. The \ac{bilstm} was then trained using the string dictionary produced by \ac{pcfg}. They compared SPRNN's performance to probability-based approaches (Markov Models and \ac{pcfg}) on both cross-site (model trained and tested on various datasets) and one-site (model trained and tested on subsets of the same dataset) scenarios. \ac{sprnn} outperforms the other two models in all circumstances, albeit it performs worse cross-site than one-site.

Based on Zhang et al.'s work\cite{zhang2018password}, the same year, Liu et al.\cite{liu2018genpass} also developed a hybrid model named GENPass that can be generalized to cross-sites attacks. The model preprocesses a password by encoding it into a series of units that are then given tags based on \ac{pcfg} (e.g., 'password123' can be separated into two units: 'D8' and 'L3'). After that, \ac{lstm} is used to create passwords. Additionally, they built a \ac{cnn} classifier to determine which wordlist the password is most likely to originate from. The results indicate that GENPass can achieve the same degree of security as the \ac{lstm} model alone in a one-site test while generating passwords with a substantially lower rank. GENPass enhanced the matching rate by 16 to 30\% when compared to \ac{lstm} alone in the cross-site test.

\subsubsection{Autoencoders}

Autoencoders are any model architecture that is composed of two submodules: an encoder and a decoder. The encoder is responsible for learning the representation of the source text at each time step and generating a latent representation of the whole source sentence, which the decoder uses as an input to build a meaningful output of the original phrase. Typically, autoencoders are employed to deal with sequential data and various NLP tasks, such as machine translation, text summarization, and question answering. \ac{rnn} and \ac{cnn} are often used as encoder and decoder components, respectively.

Pasquini et al.\cite{pasquini2021improving} applied this strategy to a dataset containing leaked passwords, using \ac{gan} and Wasserstein Autoencoders (WAEs) to develop a suitable representation of the observed password distribution rather than directly predicting it. Their methodology, called \ac{dpg}, can guess passwords that are unique to the password set. and they are the first to apply completely unsupervised representation learning to the area of password guessing.

\subsubsection{Attention-based models}

When we use the phrase "Attention" in English, we mean concentrating our focus on something and paying closer attention. The Attention mechanism in Deep Learning is based on this principle, and it prioritizes certain tokens (words, letters, and phrases) while processing text inputs. This, intuitively, aids the model in gaining a better knowledge of the textual structure (e.g., grammar, semantic meaning, word structure, and so on) and hence improve text classification, generation and interpretability. In language models, attention mechanisms are often utilized in combination with \ac{rnn} and \ac{cnn}. However, even with \ac{lstm}, these models cannot manage lengthy dependencies since transforming the whole source sentence to a fixed-length context vector is challenging when the source sentence is too long. As a result, Transformers\cite{vaswani2017attention} were invented that were built just on Attention, without convolution or recurrent layers. \ac{bert}\cite{devlin2018bert}, ELMO\cite{peters2018deep}, and GPT\cite{radford2018improving} are all well-known instances of attention-based applications built on top of Transformers.

\begin{table*}[h]
  \begin{center}
    \caption{A comparison for Deep Learning Models used for Password Guessing.}
    \label{tab:models}
    \begin{tabular}{l|l|l|l} 
      \hline
      \textbf{Category} & \textbf{Methods} & \textbf{Models used}& \textbf{Year}\\
      \hline 
      \hline
      Autoencoders & \ac{dpg}\cite{pasquini2021improving} & WAE, \ac{gan} & 2021 \\
      \ac{gan} & REDPACK\cite{nam2020generating} & \ac{iwgan}, RaGAN, HashCat, \ac{pcfg} & 2020 \\
      \ac{gan} & PassGAN\cite{hitaj2019passgan}  & \ac{iwgan} & 2019 \\
      Attention & Language Model\cite{li2019password}  & \ac{bert}, \ac{lstm} & 2019 \\
      \ac{rnn} & GENPass\cite{liu2018genpass} & \ac{pcfg}, \ac{lstm}, \ac{cnn} & 2018 \\
      \ac{rnn} & SPRNN\cite{zhang2018password} & \ac{pcfg}, \ac{bilstm} & 2018 \\
      \ac{rnn} & \ac{fla}\cite{melicher2016fast} & \ac{lstm} & 2016 \\
      \hline
    \end{tabular}
  \end{center}
\end{table*}

Li et al.\cite{li2019password} proposed a curated Deep Neural Network architecture consisted of five \ac{lstm} layers and an output layer, and then tutored and improved the created model using \ac{bert}. They proved that the tutoring process by \ac{bert} can help increase the model performance significantly.

A comparison of prior published deep learning-based password guessing tools is illustrated in Table\ref{tab:models}.

\begin{table*}[!b]
  \begin{center}
    \caption{Datasets used for training and evaluating deep learning models. Size is the number of passwords in the dataset.}
    \label{tab:datasets}
    \begin{tabular}{l|l|l}
    \hline
      \textbf{Name} & \textbf{Size} & \textbf{Brief Description} \\
      \hline
      \hline
      Yahoo\ & $4.4\times10^5$ & A web services provider.\\
      phpBB &  $3\times10^5$ & A software website.\\
      RockYou\ & $1.4\times10^7$ & A gaming platform.\\
      Myspace\ & $5.5\times10^4$ & A social networking platform.\\
      SkullSecurityComp\ & $6.7\times10^6$ & Compilations of passwords lists.\\
      LinkedIn & $1.3\times10^6$ & A social online platform.\\
      \hline
    \end{tabular}
  \end{center}
\end{table*}

\section{Methodology}

\subsection{Problem Statement}
It is a consensus in academia that password strength estimates based on entropy, and marginal guesswork analysis\cite{pliam2000incomparability} are insufficient. Rather than that, it is more suitable to quantify it in terms of the number of attempts required to find a password collision using the most efficient adversarial attacks\cite{dell2015monte}. Thus, password guessing attacks can be used for password security analysis and be developed into \ac{psm} that can be integrated into an intelligent user interface to encourage users to establish stronger passwords. 
I will concentrate my study on utilizing \ac{gan} to complete the password guessing task.

\subsection{\ac{gan}}

\begin{figure}[h]
  \includegraphics[width=0.45\textwidth]{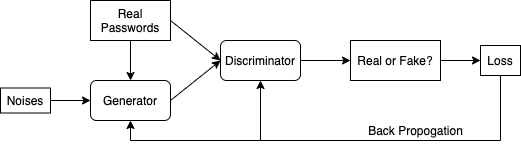}
  \caption{Model architecture diagram.}
  \label{fig:model}
\end{figure}

Unlike the previously described deep learning-based algorithms commonly employed in Natural Language Processing tasks, \ac{gan}\cite{goodfellow2014generative} has been used to construct simulations of pictures, texts, and voice across all domains. Behind the scenes, \ac{gan} consists of two sub-modules: a discriminator (D) and a generator (G), both of which are built of deep learning neural networks. G accepts noise or random features as input; learns the probability of the input's features; and generates fake data that follows the distribution of the input data. While D makes every effort to discriminate between actual samples and those created artificially by G by estimating the conditional probability of an example being false (or real) given a set of inputs (or features). The model architecture diagram is illustrated in Figure \ref{fig:model}. This cat-and-mouse game compels D to extract necessary information in training data; this information assists G in precisely replicating the original data distribution. D and G compete against one another during the training phase, which progressively improves their performance with each iteration. Typically, proper gradient descent and regularization techniques must be used to accelerate the whole process. More formally, the optimization problem solved by \ac{gan} can be summarized as follows:

\begin{equation*}\label{GAN}
  \begin{aligned}
  \min_{\theta_G}\max_{\theta_D}\sum_{i=1}^{n}logf(x_i;\theta_D) + \sum_{i=1}^{n}log(1-f(g(z_j;\theta_G);\theta_D))
  \end{aligned}
\end{equation*}
where $f(x;\theta_D)$ and $g(z_j;\theta_G)$ represents the discriminator D and the generator G respectively. 

The optimization demonstrates the min-max game between D and G. After the initial \ac{gan} work was published in 2014, several enhancements were made, and Hitaj et al.\cite{hitaj2019passgan} leveraged the \ac{iwgan}\cite{gulrajani2017improved} to apply \ac{gan} on password guessing, which is the first in literature. They trained the discriminator using a collection of leaked passwords (actual samples). Each iteration brings the generator's output closer to the distribution of genuine passwords, increasing the likelihood of matching real-world users' passwords. Consequently, PassGAN outperformed current rule-based password guessing tools and state-of-the-art machine learning password guessing technologies (\ac{fla}) after sufficient passwords were generated ($10^9$). They matched 51\% - 73\% of passwords when combining PassGAN with HashCat, compared to 17.67\% when using HashCat alone and 21\% when using PassGAN alone. One disadvantage of PassGAN is that it has intrinsic training instability due to the final softmax activation function in the generator, which may easily result in the network being vulnerable to vanishing gradients, lowering the guessing accuracy.

Following the publication of PassGAN in 2019, other researchers saw the possibilities of using \ac{gan} for password guessing, and more refinements have been done on top of PassGAN. In 2020, Nam et al.\cite{nam2020generating} developed REDPACK that employs a variant of \ac{gan} in conjunction with various password generation models for improved cracking performance. They suggested rPassGAN in their prior study, which enhanced PassGAN by altering its fundamental Neural Network architecture. More precisely, they employed \ac{rnn} in PassGAN instead of ResNet in PassGAN's original paper. However, during rPassGAN's training process, it became unstable at times, and REDPACK introduced the RaSGGAN-GP cost function to stabilize the training process. Nam et al. also introduced a selection phase to REDPACK, during which the password candidates are generated using several password generators (Hashcat, acpcfg, and rPassGAN). The discriminator then determines the chance of each generator's password candidates being realistic and sends the candidates with the greatest probability to password cracking tools such as HashCat. We regard PassGAN to be a good representation of \ac{gan}-based password guessing tools, and PassGAN enhancements will be the focus of my study.

\subsection{Dataset}
Table \ref{tab:datasets} summarizes the publicly accessible datasets that are often used in password guessing. While one disadvantage of prior work on the dataset is that most studies did not include the critical nature of password policy. In reality, most websites impose password policy restrictions to encourage users to generate strong passwords. For example, create a password with at least one special symbol and eight characters in both lowercase and uppercase\cite{bonneau2010password}. In my research, I will consider the password policy when preprocessing datasets to imitate real-world password creation circumstances more accurately. 

Dataset that I can use for training and evaluation:\href{https://weakpass.com/wordlist/1920}{Weakpass} (Contains only passwords that are suitable for password policy, length more than 8, with 2133708093 passwords) and the smaller dataset from \href{https://github.com/danielmiessler/SecLists/tree/master/Passwords/Leaked-Databases}{GitHub}.

\subsection{Experimental design}
Tensorflow or Pytorch serves as the primary framework for the experiments. Reproducing PassGAN's work\cite{hitaj2019passgan} is the first step in my research. However, there are limitations of PassGAN and to address which and improve PassGAN will be the focus of my work. The vanilla \ac{gan} stays unstable throughout the training process because the discriminator's steep gradient space, which results in mode collapse during the generator's training phase\cite{thanh2019improving}. As a consequence, the generator is prone to deceive the discriminator before mastering the art of creating more realistic passwords. Although the \ac{iwgan} utilized in PassGAN uses the Wasserstein Loss function, which, in comparison to the Binary Cross Entropy Loss used in the vanilla \ac{gan}, aids in the resolution of mode collapse and vanishing gradient difficulties in some degree. However, in order to train \ac{gan} using Wasserstein-loss, the discriminator must be 1-Lipschitz continuous, which means that the gradient's norm should be at most 1 at every point. While \ac{iwgan} used a gradient penalty to guarantee 1-Lipschitz continuity, Wu et al.\cite{GNGAN_2021_ICCV} proved that "the Lipschitz constant of a layer-wise Lipschitz constraint network is upper-bounded by any of its k-layer subnetworks." Thus, I will explore alternative techniques aimed at fulfilling the Lipschitz constraint while applying Wasserstein loss or stabilizing the training process by further regularization and normalizing on the discriminator\cite{GNGAN_2021_ICCV, gulrajani2017improved, jabbar2021survey, thanh2019improving, wei2018improving, wu2018wasserstein, terjek2019adversarial, jiang2018computation}. By comparing the matching result of the testing set to the synthesized set, we can determine if the model we build can outperform PassGAN.

\begin{table}[h]
\begin{center}
\caption{Unique password distribution in the Rockyou dataset}
\begin{tabular}{|l|l|}
\hline
\textbf{Length} & \textbf{Frequency}\\
\hline 
\hline
$<8$ & 15.55\% \\
8-10 & 38.16\% \\
10-12 & 29.32\% \\
12-16 & 14.18\% \\
$>=16$ & 2.80\% \\
Total & $1.43*10^7$ \\
\hline
\end{tabular}
\label{tab:rockyou}
\end{center}
\end{table}

The datasets listed in Table\ref{tab:datasets} will be used to train and test my curated model. Small, medium and big password candidate files will be produced to evaluate the model's guessing capability. The big one may be up to 6.5 GB in size and include over $8\times10^9$ unique passwords. To train a deep learning model and to produce such an enormous dataset requires substantial computational resources. A Tesla P100 GPU or above is required to reduce training and generation time. A system with at least 16 GB of RAM is required to prevent the process from being killed due to running out of memory while reading the created file to calculate guessing accuracy. 
\begin{figure}[h]
  \includegraphics[width=0.45\textwidth]{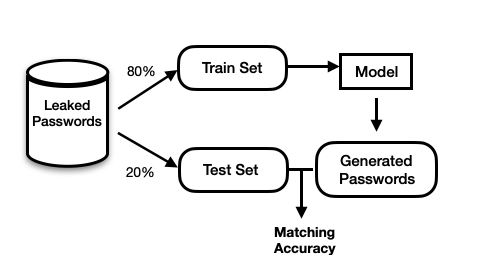} 
  \centering
  \caption{Data Flow Diagram.}
  \label{fig:data}
\end{figure} 

Note that the work is based on the assumption that the training set and testing set has the same distribution, therefore by simulating samples from the training set, the produced samples can substantially approximating the test set. The two sets are produced by dividing leaked passwords to 4:1 in portion, and no duplicate exist in the two sets. By shuffling before splitting, we assume that the two sets have the same distribution. The dataflow diagram is illustrated in Figure\ref{fig:data}.

\section{Experimental results}
I used Rockyou Dataset to train the \ac{iwgan} model, the dataset it preprocessed so that all passwords are unique, there is no password both exiting in the testing set and training set. A distribution of the passwords based on length is illustrated in Table\ref{tab:rockyou}. There are two minor modifications in my experiment and Hitaj et al. \cite{hitaj2019passgan}'s. First, when preprocessing the dataset, I restrict all passwords to 8 to 12 in length, which is more in line with the modern password creating policy. While Hitaj et al. limited all passwords to less than 10. However, in the real world scenario, a password with less than eight characters is regarded as weak and even a rule-based or probability-based password guessing tool can easily guess them. In my work, I want to demonstrate the capability of deep learning-based password guessing tools against other kinds, so the password policy restriction is considered. Second, I modified the activation function in the last layer of the Generator to Tanh instead of Softmax, and this made the model convergence faster. 

The training and testing set each have $8.44*10^6$ and $2.11*10^6$ respectively. The model was set to train for 200000 iterations as stated in the origin PassGAN's work. But after 120000 iterations, the loss of both discriminator and generator became plateau, and it took a Tesla P100 GPU 12 hours to train the dataset for 75000 iterations. By generating fake passwords using the 120000 checkpoints, compared the generated password with the testing dataset, the matching accuracy is 3.99\% when generating $9.3*10^7$ unique passwords; and when generating $10^8$ passwords. Compared with Hitaj et al's work, 

\section{Discussion}
Password Guessing with Deep Learning is a relatively new area, with the first related paper published in 2016\cite{melicher2016fast} using \ac{lstm} and first \ac{gan} related model published in 2019\cite{hitaj2019passgan}. With continuous advancements in machine learning and deep learning methods, bringing these advancements and breakthroughs to the password guessing domain is an exciting research topic. Nonetheless, several issues persist. I will introduce the open problems in password guessing and the associated future research possibilities in this section.

\subsection{Open problems}
\begin{itemize}
 \item Insufficiency of training resources. The majority of password guessing tools are trained and tested using leaked datasets shown in Table\ref{tab:datasets}. While it is straightforward to compare the performance of each model, the generalizability of the models cannot be guaranteed.
  \item The rate of successfully guessing the passwords is still low even using a combination of three types of models (rule-based, probability-based and deep learning-based). On the one-site password guessing problem, the best guessing accuracy is 51.8\% after $10^{10}$ guesses\cite{pasquini2021improving}. For cross-cite guessing, the best performance is 23\% after $10^{12}$ guesses\cite{liu2018genpass}. There is much space for improvement in the password guessing problem in terms of success rate.
  \item The distribution or pattern of passwords disclosed five years ago may change from the passwords used today, yet most suggested models train and test on the same dataset, which is impractical in the real world. However, if the training and testing sets have different distributions, the model may experience concept drift. Even Liu et al.\cite{liu2018genpass}'s hybrid model can successfully guess 23\% of passwords correctly after generating $10^{12}$ passwords. The model's capacity to generalize effectively to a different dataset than the testing set is critical because assessing cross-site is more realistic and requires further study. 
  \item The majority of research in the literature made no distinction between various password policies. However, most websites demand a mix of numeric characters, special symbols, and special characters. Prior study\cite{tan2020practical} has shown that varying password regulations, such as minimum-length and minimum-class requirements, can have a considerable influence on both usability and security. As a result, it is desirable to test password guessing tools against a variety of password policies.
\end{itemize}

\subsection{Future work}
\begin{itemize}   
\item Although the provided deep learning algorithms are appropriate for password guessing attacks; the suggested models are somewhat simplistic. With advancements in the natural language processing and the computer vision fields, \ac{bert}, variants of \ac{gan}, the self-attention mechanism\cite{choromanski2020rethinking}, and autoencoders have all seen significant improvements, and these newly invented techniques and architectures can be used to improve password guessing performance.
\item \ac{gan} can also be utilized in targeted assaults. Once the attackers get access to the users' \ac{pii}, they may condition \ac{gan}\cite{mirza2014conditional} on the precise terms associated with the users' \ac{pii}, causing the generator to pay more attention to the region of the search space containing these keywords.
\item The estimations of Deep Learning models can be used to develop trustworthy \ac{psm} that can assist users in creating passwords that are resistant to guessing attacks. The challenges are: to develop an application capable of storing and querying the compressed neural network and an interactive \ac{psm} user interface that can prompt users to build more secure passwords; convert deep learning results to interpretable rank as password strength. One solution is to store \ac{gan} generated samples by different attempts in dictionaries and search for which dictionary the user-created password belongs to. For example if the created password first exists in the dictionary with 10000 generated passwords, then the strength is assigned to 4, if it first appears in the generated $10^9$ dictionary, then the strength is assigned to 9. The higher number the more secure as it needs exponentially more computation to guess the password correctly.
\item To overcome the paucity of training and testing resources, researchers might train and test models using custom or site-specific data. For example, passwords generated during user studies to evaluate different password usability and security experiments may be utilized in our purpose.
\item When reviewing novel password guessing methods, it is necessary to consider password policies. Additionally, transfer learning may improve models' performance when there is a shortage of training data for a particular policy. For example, we may train a pre-trained model using generic passwords, freeze the network's bottom layers and retrain the model using only passwords that adhere to specific criteria. Melicher et al.\cite{melicher2016fast} employed transfer learning to train models for certain password restrictions. However, the choices are not exhaustive.   
\end{itemize}

\section{Conclusions}

In this paper, I compared the commonly deployed rule-based, probability-based password guessing tools with deep learning-based approaches from a high level, and introduced the deep learning techniques used in the password guessing area, including \ac{rnn}, autoencoder, \ac{gan} and the Attention mechanism. A research idea involving using a \ac{gan}-based model on password guessing attacks is proposed, along with the experimental methodology and resources required.

Deep learning has made tremendous strides in recent years and will continue to do so in the near future. Deep learning application scenarios abound in our daily lives. Password guessing is a novel and exciting setting that will undoubtedly benefit from deep learning's growth. I hope that with the attention of researchers, the issues raised in this work will find promising solutions.

\section{Appendix}
\begin{figure*}[!ht]
\begin{tikzpicture}[
    mindmap,
    concept color = red!30,
    every node/.style = {concept},
    grow cyclic,
    level 1/.append style = {
        concept color = red!20,
        level distance = 4.5cm,
        sibling angle = 120
    },
    level 2/.append style = {
        concept color = red!10,
        level distance = 3cm,
        sibling angle = 45
    }
]
\node{Password Guessing Attacks}
	child { node {Types}
	child { node {Online and Offline}}
	child { node {Targeted and Non-Targeted}}
	child { node {Static and Dynamic}}
	}
	child { node {Intention}
	child { node {Password Strength Meters}}
        child { node {Honeyword Generation}}
	}
	child { node {Techniques}
	child { node {Rule-based}
	child { node {HashCat\cite{hranicky2019distributed}}}
	child { node {John the Ripper}}
	}
	child { node {Probability-based}
	child { node {PCFG\cite{weir2009password}}}
	child { node {Markov Models\cite{narayanan2005fast}}}
	}
	child { node {Deep Learning-based}
	child { node {RNN/LSTM
	\cite{melicher2016fast, zhang2018password,liu2018genpass}}}
	child { node {Autoencoder
	\cite{nam2020generating}}}
	child { node {GANs
	\cite{hitaj2019passgan, pasquini2021improving}}}
	child { node {Attention-Mechanism
	\cite{li2019password}}}
	}
	};
\end{tikzpicture}
\caption{A mind map for password guessing attacks} \label{fig:mindmap}
\end{figure*}
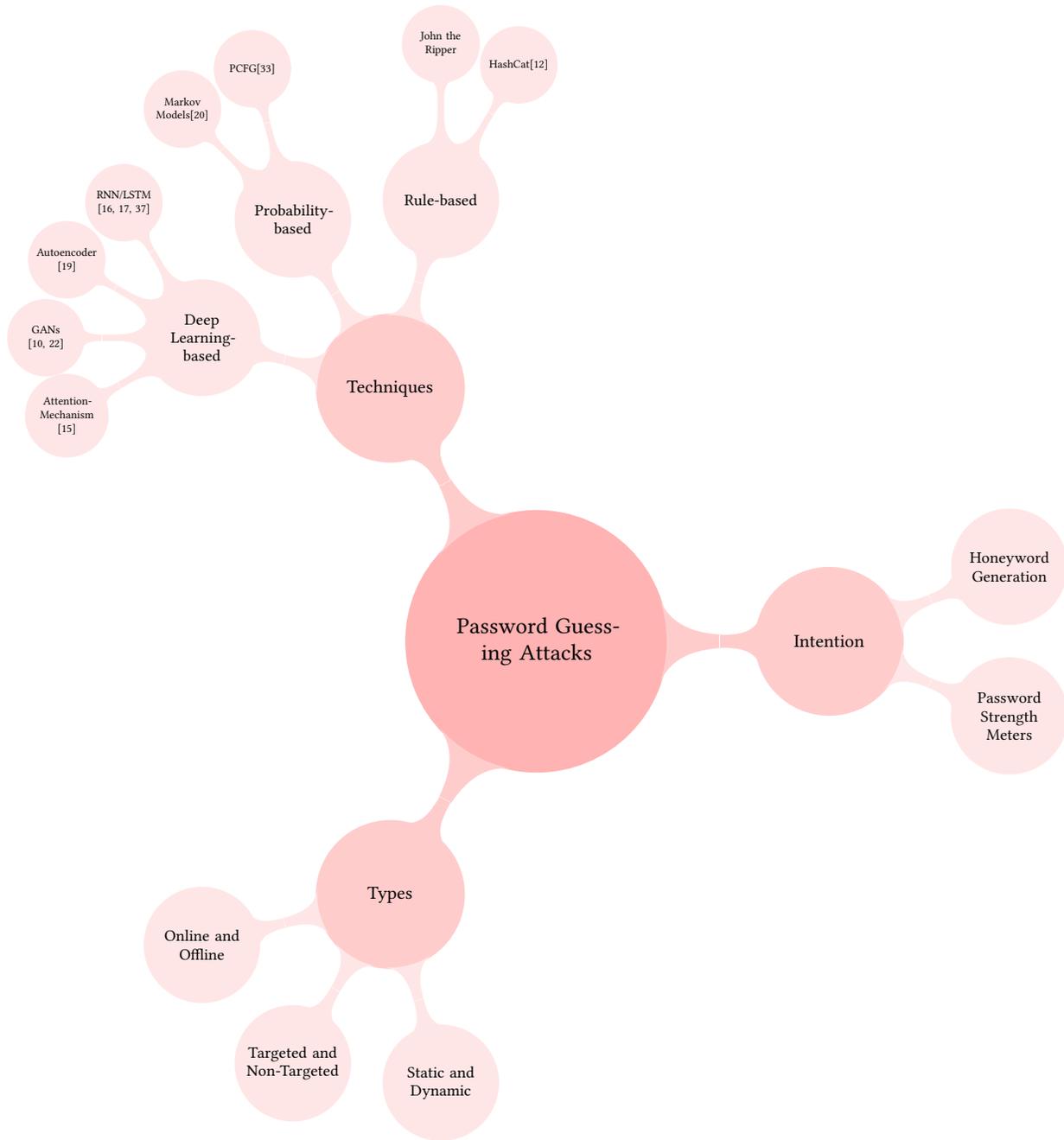

Please see Figure\ref{fig:mindmap} for the mind map for the password guessing attacks.

\printglossaries
\bibliographystyle{acm}
\bibliography{References}
\end{document}